# Constraints on CDM theories from observations of massive, X-ray luminous clusters of galaxies at high redshift

G. A. Luppino and I. M. Gioia[1]

Institute for Astronomy, University of Hawaii
2680 Woodlawn Drive, Honolulu, HI USA 96822
email: ger@hokupa.ifa.hawaii.edu, gioia@galileo.ifa.hawaii.edu

## Abstract

During the course of a gravitational lensing survey of distant, X-ray selected, EMSS clusters of galaxies, we have studied 6 X-ray luminous ($L_x > 5 \times 10^{44}~h_{50}^{-2}$ erg s$^{-1}$) clusters at redshifts exceeding $z = 0.5$. All of these clusters are apparently massive. In addition to their high X-ray luminosity, two of the clusters at $z \sim 0.6$ exhibit gravitationally lensed arcs. Furthermore, the highest redshift cluster in our sample, MS 1054−0321 at $z = 0.826$, is both extremely X-ray luminous ($L_{0.3-3.5\,\mathrm{keV}} = 9.3 \times 10^{44}~h_{50}^{-2}$ erg s$^{-1}$) and exceedingly rich with an optical richness comparable to an Abell Richness Class 4 cluster.

In this *Letter*, we discuss the cosmological implications of the very existence of these clusters for hierarchical stucture formation theories such as standard $\Omega = 1$ CDM, hybrid $\Omega = 1$ C+HDM, and flat, low-density $\Lambda$ + CDM models.

*Subject headings:*  cosmology: observations — large-scale structure of the universe — gravitational lensing — galaxies: clusters: individual (MS 1054−0321) — X-rays: galaxies

## 1. Introduction

The very existence of massive clusters at high redshift ($z > 0.5$) is problematic for standard cold dark matter (CDM) theories of hierarchical structure formation. The CDM model, as originally formulated, assumed that the large scale distribution of galaxies could be used to trace the underlying mass. It became clear early on, however, that observed structure and galaxy velocities on small scales demanded a biased model of galaxy formation where the galaxy number fluctuations $\delta N/N$ are more pronounced than the actual mass fluctuations $\delta M/M$ by some biasing factor $b > 1$ (Davis & Peebles 1983). This bias factor is defined as the ratio of the galaxy to mass fluctuations within an $8\,h^{-1}$ Mpc radius sphere. Since the field galaxy number fluctuations on this scale are observed to be close to one (Davis & Peebles 1983), the bias parameter is often expressed as $b = 1/\sigma_8$, the inverse of the rms density fluctuation spectrum filtered on this $8\,h^{-1}$ Mpc scale. Large clusters of galaxies are rare objects that formed from exceedingly large mass

---

[1] Also from Istituto di Radioastronomia del CNR, Via Gobetti 101, I-40129, Bologna, ITALY

density perturbations. Consequently, the abundance of clusters as a function of mass provides a sensitive, and nearly model free way to measure the CDM spectrum normalization. A factor of two change in $\sigma_8$ alters the expected number density of clusters by two orders of magnitude (Carlberg et al. 1993; Bahcall & Cen 1992; White et al. 1993). Furthermore, the change in the comoving number density of rich clusters as a function of redshift provides another important observational constraint on the normalization of the CDM spectrum. The number density of clusters evolves faster in a high bias CDM universe than in a low bias CDM universe (Cen & Ostriker 1994).

Various groups have attempted to measure and analyze the abundance of rich clusters of galaxies at both high and low redshift. Bahcall & Cen (1992) find that cluster abundances at low-$z$ are inconsistent with $\Omega_0 = 1$ CDM for any value of $\sigma_8$, and are best fitted by $\sigma_8 \sim 1$, low density ($\Omega_0 \sim 0.25$) CDM models, either with or without a cosmological constant. Conversely, White et al. (1993) argue that observed low-$z$ cluster abundances suggest $\sigma_8 \sim 0.6$ for an $\Omega_0 = 1$ universe. At higher redshift, cluster abundances have been more difficult to determine, but results seem to indicate an overabundance of massive clusters, again in favor of a low, or even unbiased CDM model with $\sigma_8 \sim b \sim 1$. For example, Evrard (1989) argued that the existence of only three clusters with high velocity dispersions in the intermediate redshift range of $z \sim 0.4 - 0.6$ is difficult to reconcile with a high value for $b$. Peebles, Daly & Juszkiewicz (1989) showed that the existence of *just one* cluster of $\sim 10^{15}\ M_\odot$ in the redshift range $z \sim 0.5 - 0.6$ (e.g. CL 0016+16) implied an upper limit of $b < 1.6$. Their limit drops to $b < 1.38$ if just one Coma-like cluster is found between redshifts $z \sim 0.7 - 0.8$. And such clusters do appear to exist. Gunn (1990) lists 4 clusters in the range $0.7 < z < 0.85$ and one at $z \sim 0.9$, two of which have measured velocity dispersions of $\sim 1000$ km s$^{-1}$ (i.e. comparable to Coma).

Of course the statistics on the abundance of high-$z$ clusters are poor because the presently-known number of such clusters is so small. Furthermore, one might argue that many of these clusters may not be as massive as they appear due to inflation of the measured velocity dispersion from field galaxy contaminaton and projection effects (Frenk et al. 1990). Recent optical and X-ray observations of some optically-selected clusters with $z > 0.6$ may lend support to these contamination arguments. Thimm & Belloni (1994) presented observations of two of the highest redshift Gunn, Hoessel & Oke (1986) clusters and found that one of them (CL 2155+0334) was not a cluster at all, but a chance superposition of field galaxies. Castander et al. (1994) observed 10 distant, optically-selected clusters (from Gunn 1990) with *ROSAT*. Only 5 of their clusters had known redshifts, but all of those 5 were at $z \gtrsim 0.7$. Castander et al. detected X-ray emission from only 2 (and possibly a third) of these clusters. After converting their detected fluxes to luminosities in the standard way using a $T \sim 7$ keV thermal spectrum and correcting for the extended emission assuming a $\beta = 2/3$ model for the cluster density profile, Castander et al. found the detected clusters were relatively modest X-ray emitters with much lower X-ray luminosities ($L_{0.1-2.4\,\mathrm{keV}} \sim 1 \times 10^{44}\ h_{50}^{-2}$ erg s$^{-1}$) than equivalently-rich low or intermediate redshift clusters. In a similar study, Nichol et al. (1994) conducted a *ROSAT* search for distant X-ray clusters in fields containing either previously-identified, optically-selected clusters or wide angle radio sources. Two of their objects were detected with *ROSAT*; one corresponded to one of their optically-selected clusters ($z$=0.61), while the other corresponded to a radio-selected cluster estimated (from photometry) to be at $z \sim 0.7$. Again, after converting the detected fluxes to luminosities in approximately the same manner as Castander et al. , Nichol et al. found that both of their clusters were also relatively modest X-ray emitters ($L_x \sim 10^{44}\ h_{50}^{-2}$ erg s$^{-1}$). Nichol et al. estimated a lower limit to the comoving number density of such clusters ($n_z > 1.2 \times 10^{-7}\ h_{50}^3$ Mpc$^{-3}$) and argued this was marginally consistent with a standard $b \sim 2$ CDM universe.

While these data would seem to imply that extremely rich systems with large amounts of X-ray emitting gas do not exist at these high redshifts, we will show several examples that indicate this is not the case. In this paper, we present preliminary data on 6 rich, massive, X-ray luminous clusters at $z > 0.5$ ($\overline{z} = 0.66$, $z_{\mathrm{max}} = 0.826$). The clusters were selected from the *Einstein Observatory* Extended Medium Sensitivity Survey (EMSS; Gioia et al. 1990a) and are the most distant members of a sample of 40 such clusters we are examining for evidence of gravitational lensing (Luppino & Gioia 1992; Luppino et al. 1995). In the following section, we briefly discuss the EMSS cluster sample, and describe some of the properties of the six $z > 0.5$ clusters. In §3, we discuss the implications the very existence of these clusters have for various CDM models. Unless specifically stated otherwise, all relevent cosmological quantities were computed assuming $H_0 = 50$ and $\Omega = 1$, however we will often list quantities in terms of $h_{50} = H_0/50$ to show the explicit dependence of that quantity on the value of the Hubble parameter.

## 2. Highest redshift EMSS clusters

The EMSS cluster sample consists of 103 clusters of galaxies extracted from a total of 835 EMSS sources. The most recent optical, X-ray and radio data for these clusters have been compiled and presented by Gioia & Luppino (1994). The cluster subsample chosen for our gravitational lensing survey (Luppino et al. 1995) was subjected to the following selection criteria: 1) the fluxes of the sources had to exceed $1.33 \times 10^{-13}$ erg cm$^{-2}$ s$^{-1}$, 2) the sources had to lie North of declination $\delta = -40^\circ$, and 3) the clusters had to be distant ($z > 0.15$), and 4) X-ray Luminous ($L_x > 2.0 \times 10^{44}\ h_{50}^{-2}$ erg s$^{-1}$). Note, X-ray luminosities were determined in the conventional way assuming a $T \sim 6$ keV thermal spectrum and correcting for the extended emission assuming a $\beta = 2/3$ model for the cluster surface brightness (the reader should refer to Gioia & Luppino 1994 and references therein for the details of this procedure). These selection criteria resulted in 40 clusters of galaxies spanning a redshift range of $0.15 \leq z < 0.83$, and included all of the EMSS clusters with $z > 0.5$.

Optical imaging observations of these clusters were carried out using the UH 2.2 m, with some data obtained using the CFHT and Keck telescopes. The details of the imaging observations and data reduction can be found in Luppino et al. (1995). The majority of our spectra were obtained using the MOS on the CFHT or LRIS on Keck. Details of the spectroscopic observations and data reduction will be presented in a later paper.

Table 1. $z > 0.5$ EMSS clusters

| (1) | (2) | (3) | (4) | (5) |
|---|---|---|---|---|
| Cluster | $z$ | $L_x$(0.3–3.5 keV) $10^{44}\ h_{50}^{-2}$ erg s$^{-1}$ | $N_{0.5}$ | $\sigma$ km s$^{-1}$ |
| MS0015.9+1609 | 0.546 | 14.64 | 66 ± 6 | 1324[b] |
| (or CL 0016+16) | | | 46[a] | ... |
| MS0451.6−0305 | 0.55 | 19.98 | 47 ± 5 | ... |
| MS1054.5−0321 | 0.826 | 9.28 | 50 ± 10 | $1643^{+806}_{-343}$ |
| MS1137.5+6625 | 0.782 | 7.56 | 30 ± 5 | ... |
| MS1610.4+6616 | 0.65 | 6.36 | ... | ... |
| MS2053.7−0449 | 0.583 | 5.78 | 20 ± 5 | ... |

[a] Koo 1981; [b] Dressler & Gunn 1992

In Table 1, we list the six $z > 0.5$ EMSS clusters. All 6 of these clusters are extremely X-ray luminous ($L_x \gtrsim 5 \times 10^{44}\ h_{50}^{-2}$ erg s$^{-1}$). Two, in fact, are among the most X-ray luminous clusters known—at any redshift. In the clusters in this paper, we are reasonably confident that the X-ray emission comes from the cluster and not from an AGN in the field. Only 2 of the 6 cluster fields have a radio source detected at 6 cm with the VLA, and in both cases that source is coincident with the brightest cluster galaxy (see Gioia & Luppino 1994). Thus any AGN in these fields would have to be radio quiet. Furthermore, during the course of the EMSS optical identifications (Stocke et al. 1991), any blue, stellar objects in the fields were observed spectroscopically. While there are several cases in the EMSS where AGN are present in clusters, and the identification is not certain, this is not the case for the six clusters discussed in this paper.

In addition to high X-ray luminosity, there is other compelling evidence that these clusters are genuinely massive. Two of the clusters at $z \sim 0.6$ contain large lensed arcs that allow us to make crude, but independent estimates of the projected mass in the clusters cores (Luppino et al. 1995). We find central masses of $\sim 5 \times 10^{14}\ M_\odot$ enclosed within a $237 h_{50}^{-1}$ kpc critical line for MS 0451−0305 ($z$=0.55) and $\sim 1.3 \times 10^{14}\ M_\odot$ enclosed within $121 h_{50}^{-1}$ kpc for MS 2053−0449 ($z$=0.583). In both cases, we assume the source redshift is twice that of the cluster lens, but these masses drop by only 30% if the source is moved to $z_s \sim 2$. We can also estimate the cluster central velocity dispersion from the lensing geometry if we make the simple assumption that the cluster potentials are isothermal spheres. In this case, the central velocity dispersion ($\sigma$) is related to the lensing factor ($D_s/D_l\, D_{ls}$ where $D_s$, $D_l$, and $D_{ls}$ are the angular diameter distances to the source, the lens and from the lens to the source, respectively), and distance of the arc to the center of the lens ($r_c$) by $\sigma^2 = (c^2/4\pi)(D_s/D_l\, D_{ls})r_c$. Again assuming the source redshifts are twice the lens redshifts, we find central velocity dispersions of $\sim 1750$ km s$^{-1}$ and $\sim 1200$ km s$^{-1}$ for MS 0451−0305 and MS 2053−0449 respectively. Note that the masses estimated above are only for the central "enclosed" cores. The total masses of each of these two clusters must surely be comparable to or greater than that of Coma.

Moving even further out in redshift, we find the two most distant EMSS clusters, MS 1137+6625 at $z = 0.782$ and MS 1054−0321 at $z = 0.826$, are also extremely X-ray luminous ($L_x \sim 10^{45}$ erg s$^{-1}$), with X-ray luminosities nearly an order of magnitude brighter than the optically-selected $z \sim 0.8$ clusters observed by Castander et al. (1994) or the optical/radio-selected $z \sim 0.6$ clusters observed by Nichol et al. (1994). Although neither of these $z \sim 0.8$ EMSS clusters show any obvious lensed arcs, their optical appearance is striking. MS 1137+6625 is rich and compact (see Fig 1 [Plate 1]), and MS 1054−0321 in particular is exceedingly rich (see Figs 2a and 2b [Plates 2 and 3]), with a projected spatial distribution of galaxies comparable to the richest low and intermediate redshift clusters such as CL 0016+16 (MS 0015.9+1609) or CL 0939+4713 (Gunn et al. 1986; Dressler et al. 1994). We measure a central richness (Bahcall 1981) for MS 1054−0321 of $N_{0.5} = 50 \pm 10$ (after background subtraction from the periphery of the CCD frame), which is significantly richer than Coma ($N_{0.5}^{\rm coma} = 28$; Bahcall 1981) and corresponds roughly to an Abell richness class 4 cluster. The $N_{0.5}$–$\sigma$ correlation (Bahcall 1981) suggests MS 1054−0321 has a velocity dispersion of $\sim 2000$ km s$^{-1}$. We have measured a *preliminary* value for the velocity dispersion (from spectra of 7 galaxies) of $1643^{+806}_{-344}$ km s$^{-1}$.

## 3. Discussion

In the previous section, we demonstrated that several rich, massive and X-ray luminous clusters do exist at moderately high redshifts. In order to obtain a quantitative estimate of the comoving number density of these clusters, we take the cosmological expression for the comoving volume element (Mathez et al. 1991),

$$\frac{dV}{dz\, d\Omega} = \left(\frac{c}{H_0}\right)^3 \frac{(q_0 z + (q_0 - 1)(\sqrt{1 + 2q_0 z} - 1))^2}{(1+z)^3\, q_0^4\, \sqrt{1 + 2q_0 z}} \qquad (1)$$

and integrate it over the appropriate solid angle and redshift range of interest. Actually performing this integral in the case of the EMSS is a complicated process. We use the $1/V_{\rm max}$ technique as applied by Gioia et al. (1990b) and Henry et al. (1992) in their investigation of the EMSS cluster X-ray luminosity function. First we consider the maximum redshift, $z_{\rm max}$, at which a cluster could have been seen for a given solid angle that has some associated limiting sensitivity. The search volume for a given cluster is then the sum of all volumes within a redshift shell that is bounded on the low end by $z_1$ and on the high end by the lesser of $z_2$ or $z_{\rm max}$ for each different sensitivity limit. The total volume for $N$ clusters over the redshift shell $z_1 \rightarrow z_2$ is then obtained by summing the volumes for each cluster

$$V_{\rm tot} = \sum_{i=1}^{N} \left( \sum_{f_{\rm lim}} 4\, d\Omega_i(f_{\rm lim}) \left(\frac{c}{H_0}\right)^3 \int_{z_1}^{z_2(f_{\rm lim}, L_x^i)} \frac{(1 + z - \sqrt{1+z})^2}{(1+z)^3\, \sqrt{1+z}}\, dz \right) \qquad (2)$$

where we have set $q_0 = \frac{1}{2}$, where $z_2$ is a function of the cluster luminosity and the appropriate survey flux limit, and where $d\Omega_i(f_{\rm lim})$ is the flux limit dependent sky coverage which ranged from 0.01 steradian for

FIGURE 1. CCD image of MS 1137+6625 at $z$=0.782. This image is an 800×800 subarray from a 3000 s $R$ exposure and measures $1.4\,h_{50}^{-1}$Mpc × $1.4\,h_{50}^{-1}$Mpc at the cluster redshift.

FIGURE 2a. CCD image of the center of MS 1054−0321 at $z$=0.826. This image is an 1024×1024 subarray from a 14400 s $I$ exposure and measures $1.9\,h_{50}^{-1}$Mpc × $1.9\,h_{50}^{-1}$Mpc at the cluster redshift.

FIGURE 2b. Wide field CCD image of the field containing MS 1054−0321. The image is a full 2048×2048 CCD frame and covers a field of view of $7'.5 \times 7'.5$ or $3.7\,h_{50}^{-1}$Mpc $\times$ $3.7\,h_{50}^{-1}$Mpc at the cluster redshift.

$f_{\text{lim}} = 1.33 \times 10^{-13}$ erg cm$^{-2}$ s$^{-1}$ to 0.22 steradian for $f_{\text{lim}} = 3.57 \times 10^{-12}$ erg cm$^{-2}$ s$^{-1}$ (see Table 3 in Henry et al. 1992). We find a volume of $5.7 \times 10^8 \, h_{50}^{-3}$ Mpc$^3$ for the redshift shell spanned by the 6 most distant EMSS clusters ($z=0.546 \rightarrow 0.826$; $\overline{z} = 0.66$) indicating a comoving number density at $\overline{z} = 0.66$ of $n_z(L_x > 5 \times 10^{44}$ erg s$^{-1}) = 1.1 \pm 0.43 \times 10^{-8} \, h_{50}^3$ Mpc$^{-3}$. Although this number density is lower than the density derived by Nichol et al., we point out that the number density we find is for clusters that are 5 to 10$\times$ more X-ray luminous. Moreover, we consider $1 \times 10^{-8} h_{50}^3$ Mpc$^{-3}$ to be a conservative lower limit for $n_z(L_x > 10^{44})$, particularly since, at $z > 0.5$, the flux-limited EMSS cluster sample will only include the most X-ray luminous clusters, and will miss the $\sim 10^{44}$ erg s$^{-1}$ clusters that no doubt exist, but whose flux would have fallen below the $1.33 \times 10^{-13}$ erg cm$^{-2}$ s$^{-1}$ EMSS detection criterion. If we restrict ourselves to just the two highest redshift EMSS clusters at $\overline{z} = 0.8$, we find a number density $n_z(L_x > 7.5 \times 10^{44}$ erg s$^{-1}) = 7 \pm 5 \times 10^{-9} \, h_{50}^3$ Mpc$^{-3}$. We remind the reader, with some trepidation, that these number densities are based on only a handful of clusters. Unfortunately, this is an unavoidable limitation of our present sample, and the large error bars reflect the small number of clusters used. In Figure 3, we plot the two densitites we find at $\overline{z} = 0.66$ and $\overline{z} = 0.8$ along with the three $z \sim 0.8$ clusters of Castander et al. (1994). On this plot we also overlay the integrated power law luminosity function for the largest redshift shell from the EMSS study of the X-ray luminosity function (Henry et al. 1992). It appears that the X-ray luminosity function at these high redshifts is consistent with the LF for EMSS clusters in the $0.30 \leq z \leq 0.60$ ($\overline{z} = 0.33$) shell.

To compare these number densities to the predicted abundances of clusters for CDM models, we take the $z \sim 0$ mass functions of White et al. (1993) or of Bahcall & Cen (1992), and shift them to high-$z$ for various values of $\sigma_8$ following Cen & Ostriker (1994). We argue that all of our $z > 0.5$ EMSS clusters are Coma-like, and thus are of order $10^{15} \, M_\odot$ within the central 1 Mpc. As we stated earlier, we believe this argument is valid given the estimated core lensing masses in two cases, and the extremely high $L_x$ in the two additional, highest-$z$ cases. We find the number density of clusters predicted by a $b \sim 2$ CDM model is far lower than the density observed, while the number density for an unbiased CDM model is roughly consistent with our value. The *COBE* measurements of CMBR temperature anisotropies (Smoot et al. 1992) also imply that $\sigma_8 = 1.08 \pm 0.25$ (Efstathiou, Bond & White 1992) and favor an unbiased version of standard CDM. The question remains, then, how to reconcile the apparent discrepancy between these observations, and observations of small scale structure that require a biased ($b \sim 2$) version of CDM.

The currently popular opinion is that standard biased CDM does not quite work (Ostriker 1993; Primack 1994), and various groups have started to investigate models with an additional free parameter; for example CDM models with a non-zero cosmological constant that are spatially flat ($\Lambda_0/3H_0^2 = 1 - \Omega_0$) and thus compatible with inflation (so-called LCDM or $\Lambda$+CDM models; Efstathiou et al. 1990; Kofman et al. 1993; Efstathiou 1994), or $\Omega = 1$ hybrid or mixed dark matter models with varying amounts of both cold and hot dark matter (C+HDM ), and with the hot dark matter (HDM) usually assumed to consist of one flavor of massive neutrino (Davis, Summers & Schlegel 1992; Taylor & Rowan-Robinson 1992). One popular variation of C+HDM that appears to agree with the *COBE* measurements, and measurements of large scale structure and large scale bulk flows assumes an $\Omega=1$ universe with $\sim 70\%$ CDM and $\sim 30\%$ HDM.

But while these popular mixed dark matter models have additional large scale power afforded by an HDM component that may help to resolve many of the problems of standard CDM, $\Omega = 1$ versions of these hybrid models may be just as vulnerable as standard CDM to the existence of massive clusters of galaxies at high redshift. For example, using N-body simulations, Jing & Fang (1994) have calculated the predicted abundance of massive clusters at intermediate and high redshifts in such a hybrid model composed of 60% CDM ($\Omega_{\text{CDM}}=0.6$), 30% HDM ($\Omega_{\text{HDM}}=0.3$), and 10% baryons ($\Omega_b=0.1$), and compared it to the predicted abundance for a flat, low-density CDM model with a non-zero cosmological constant ($\Omega_0=\Omega_{\text{CDM}}=0.3$, $h=0.75$, and $\lambda_0 = \Lambda/3 H_0^2 = 0.7$). While both models successfully account for the abundance of clusters at low redshift, and agree with the large scale structure, large scale bulk flows, and *COBE* measurements mentioned above, Jing & Fang show that the comoving number density of rich (Coma-like) clusters should evolve rapidly in the hybrid model, decreasing by an order of magnitude from $z=0 \rightarrow 0.5$ while remaining more or less constant for the $\Lambda$ + CDM model. By $z \sim 0.8$, the density of rich clusters in a C+HDM universe should be another factor of 10 lower than at $z \sim 0.5$, and thus a factor of 100 lower that the density at $z=0$. Although the five $z \sim 0.8$ clusters of Gunn (1990) appear to be in severe conflict with this form of hybrid dark matter model, Jing & Fang argue that the density of EMSS clusters at $z \sim 0.5$ (Henry et al. 1992) seems to be consistent with the predictions of C+HDM . One can therefore invoke the usual argument that the measured velocity dispersions of the $z \sim 0.8$ optically-selected clusters may somehow be inflated by contamination and projection effects. We note here, however, that the EMSS cluster density as derived by Jing & Fang from the data of Henry et al. (1992) did not include the four new $z > 0.5$ EMSS clusters. Moreover, the number density we find at $z \sim 0.8$ of $n_z \sim 7 \times 10^{-9} h_{50}^3$ Mpc$^{-3}$ for the two extremely luminous X-ray clusters having $L_x > 7.5 \times 10^{44}$ erg s$^{-1}$ may be only marginally consistent with the theory. Discovery of additional clusters of this type at these high redshifts will surely present a serious challenge to this variety of mixed dark matter model.

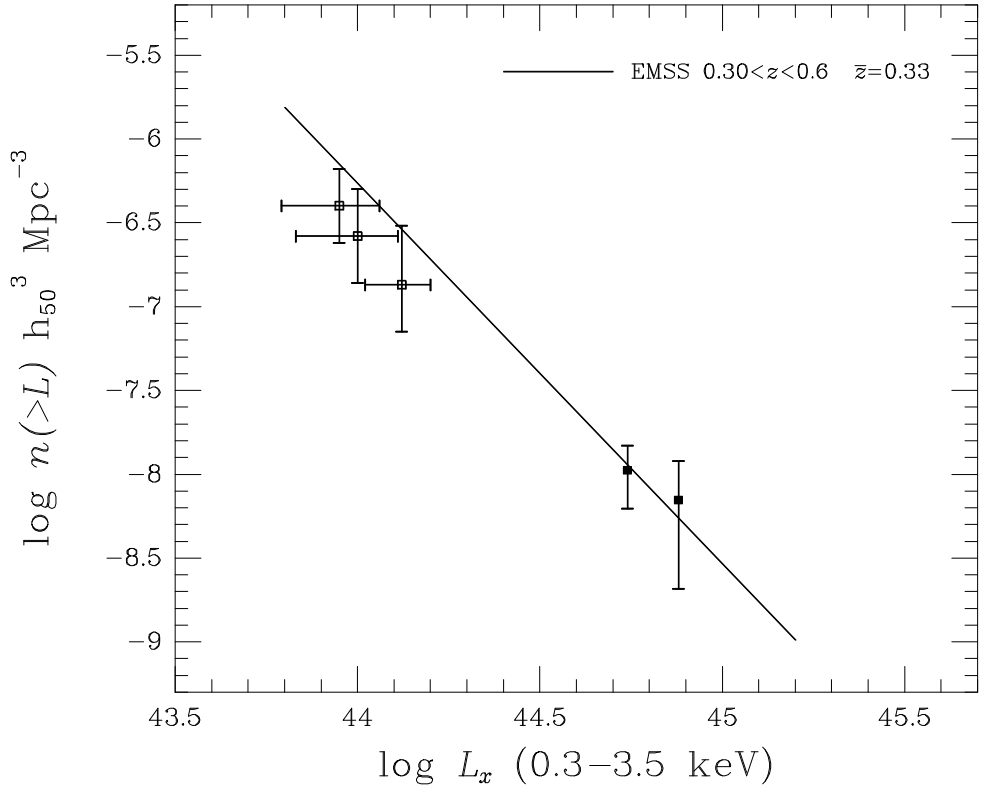

FIGURE 3. Plot of the comoving number density of distant X-ray clusters as a function of X-ray luminosity. The three points to the upper left are from Castander et al. (1994) for three $z \sim 0.8$ optically selected clusters observed with *ROSAT*. The $0.1 - 2.5$ keV *ROSAT* luminosities were converted to $0.3 - 3.5$ keV EMSS luminosities assuming $T \sim 6$ keV. The two points to the lower right are for the distant EMSS clusters discussed in this paper. The upper point is for all six $z > 0.5$ EMSS clusters ($\bar{z} = 0.66$) and the lower point is for the two $\bar{z} = 0.8$ EMSS clusters. For comparison, we also plot the integrated power law luminosity function determined by Henry et al. (1992) for their highest redshift shell ($0.30 \leq z \leq 0.60$, $\bar{z} = 0.33$) of the EMSS clusters.

On the other hand, it appears the existence of these high-$z$ clusters does not present a problem for flat, low density $\Lambda$ + CDM models, but these models lack a theoretical explanation for a non-zero vacuum energy density (see discussion in Kofman et al. 1993). Nevertheless, we also note that a non-zero cosmological constant is one of the few ways to adjust the age of a flat universe so it exceeds the lower bound on globular cluster ages when most modern distance scale methods seem to indicate $H_0 \gtrsim 75$ (Jacoby et al. 1992; Fukugita, Hogan & Peebles 1993; Pierce et al. 1994; Freedman et al. 1994).

**Acknowledgements** We are grateful to our referee, Ruth Daly, for helpful comments and recommendations. We thank Doug Clowe who helped with some of the FOCAS analysis of these clusters, and Pat Henry and Anna Wolter who verified calculations and provided helpful comments and suggestions. We are also grateful to those who read and commented on early drafts of this manuscript, including Neal Trentham, Julianne Dalcanton and Lev Kofman. Thanks to Erica Ellingson, who independently analyzed our CFHT MOS spectra of MS 1054−0321 and confirmed our redshifts. Finally, we acknowledge our collaborators on the EMSS lensing survey: J. Annis, F. Hammer and O. Le Fèvre . This work received partial financial support from NSF grant AST91-19216 and from NASA grants NAG5-1752, NAG5-1880, and NAG5-2594.